\documentclass[]{elsart}
\usepackage{epsfig}

\def\Journal#1#2#3#4{{#1} {#2} (#4) #3}

\def\NIM{Nucl. Instrum. Methods}
\def\NIMA{{Nucl. Instrum. Methods} A}

\def\NPB{{Nucl. Phys.} B}
\def\PLB{{Phys. Lett.}  B}
\def\PRL{Phys. Rev. Lett.}
\def\PRD{{Phys. Rev.} D}

\begin{document}

\begin{frontmatter}

\title{Successive Measurements of Cosmic-Ray Antiproton Spectrum in a Positive
Phase of the Solar Cycle}
\author[KOB]{T.\thinspace Maeno\thanksref{corres}},
\thanks[corres]{Corresponding author.\\ {\it E-mail address:} 
maeno@phys.sci.kobe-u.ac.jp (T. Maeno)}
\author[TOK]{S.\thinspace Orito},
\author[TOK]{H.\thinspace Matsunaga\thanksref{tukuba}},
\thanks[tukuba]{present address: University of Tsukuba,
Tsukuba, Ibaraki 305-8571, Japan} 
\author[TOK]{K.\thinspace Abe},
\author[TOK]{K.\thinspace Anraku},
\author[TOK]{Y.\thinspace Asaoka},
\author[TOK]{M.\thinspace Fujikawa},
\author[TOK]{M.\thinspace Imori},
\author[KEK]{Y.\thinspace Makida},
\author[TOK]{N.\thinspace Matsui},
\author[TOK]{H.\thinspace Matsumoto},
\author[NAS]{J.\thinspace Mitchell},
\author[KOB]{T.\thinspace Mitsui\thanksref{tohoku}},
\thanks[tohoku]{present address: Tohoku University,
Sendai, Miyagi 980-8577, Japan}
\author[NAS]{A.\thinspace Moiseev},
\author[TOK]{M.\thinspace Motoki\thanksref{tohoku}},
\author[TOK]{J.\thinspace Nishimura},
\author[KOB]{M.\thinspace Nozaki},
\author[NAS]{J.\thinspace Ormes},
\author[TOK]{T.\thinspace Saeki},
\author[TOK]{T.\thinspace Sanuki},
\author[KEK]{M.\thinspace Sasaki},
\author[MAR]{E.\thinspace S.\thinspace Seo},
\author[TOK]{Y.\thinspace Shikaze},
\author[TOK]{T.\thinspace Sonoda},
\author[NAS]{R.\thinspace Streitmatter},
\author[KEK]{J.\thinspace Suzuki},
\author[KEK]{K.\thinspace Tanaka},
\author[TOK]{I.\thinspace Ueda},
\author[MAR]{J.\thinspace Z.\thinspace Wang},
\author[ISA]{N.\thinspace Yajima},
\author[ISA]{T.\thinspace Yamagami},
\author[KEK]{A.\thinspace Yamamoto},
\author[KEK]{T.\thinspace Yoshida},
\author[TOK]{K.\thinspace Yoshimura}

\address[KOB]{Kobe University, Kobe, Hyogo 657-8501, Japan}
\address[TOK]{University of Tokyo, Tokyo 113-0033, Japan}
\address[KEK]{High Energy Accelerator Research Organization (KEK),
Tsukuba, Ibaraki 305-0801, Japan}
\address[NAS]{National Aeronautics and Space Administration,
Goddard Space Flight Center, Greenbelt,MD 20771, USA}
\address[MAR]{University of Maryland, College Park, MD 20742, USA}
\address[ISA]{The Institute of Space and Astronautical Science (ISAS),
Sagamihara, Kanagawa 229-8510, Japan}

\begin{abstract}
The energy spectrum of cosmic-ray antiprotons ($\bar{p}$'s)
has been measured by BESS successively in 1993, 1995, 1997 and 1998.
In total, 848 $\bar{p}$'s were clearly identified
in energy range 0.18 to 4.20 GeV.
From these successive measurements of the $\bar{p}$ spectrum
at various solar activity,
we discuss about the effect of the solar modulation
and the origin of cosmic-ray $\bar{p}$'s.
Measured $\bar{p}/p$ ratios were nearly identical during this perfiod, 
and were consistent
with a prediction taking the charge dependent solar 
modulation into account.

\keyword Cosmic-ray, Antiproton, Measurements, Solar modulation}
\PACS 98.70.Sa, 95.85.Ry
\end{abstract}

\end{frontmatter}

\newpage

\section{Introduction}

The origin of cosmic-ray antiprotons ($\bar{p}$'s) is
still holding a fundamental question even
after their spectrum has been measured precisely in our previous work 
\cite{BESS97}. 
In recent solar-minimum period, we have collected 458 $\bar{p}$'s 
in the energy range
between 0.18 and 3.56 GeV and observed a clear peak in
the spectrum around 2 GeV, a generic feature of ``secondary'' $\bar{p}$'s
which are produced by the interaction
of Galactic high-energy cosmic-rays with the interstellar medium. It
has become evident that the dominant component of cosmic-ray $\bar{p}$'s 
is the secondary $\bar{p}$'s.
While the energy spectrum and absolute flux are well
reproduced by several theoretical calculations \cite{BERG,BIEB}, 
there remains some diversity among those calculations
in the low energy region.
They seem
to arise from the different treatment of tertiary interactions of $\bar{p}$'s
with the interstellar medium, $\bar{p}$ production mechanism such as nuclear
sub-threshold effects, galactic propagation and solar modulations.
Moreover, although the data have still large statistical errors 
at low energies,
the observed flux appears to be higher than those calculations
in the low energy region below 1 GeV where the
secondary flux decreases sharply.
Therefore, we can not rule out the primary origins,
such as annihilation of neutralinos \cite{BERG,SILK,MI96} 
or evaporation of primordial black holes \cite{HA75,MA96}.

The cosmic-ray particles are decelerated in entering the heliosphere by the
expanding solar magnetic disturbances, and their energy spectra
observed at 1 A.U. should be deformed.
The flux
variation is expected to be small for secondary $\bar{p}$'s at low energies
because of their relatively hard spectrum, while any
softer component would exhibits a large variation. It has been pointed out
that additional soft component from a novel source can be singled out by
measuring the $\bar{p}$ spectra at different solar periods \cite{MI96}.
It has been recently suggested that the solar
magnetic polarity and the charge dependent effect cannot be neglected and
the modulation follows a 22-year cycle rather 
than an 11-year cycle \cite{BIEB}.
Actually, measurements of $e^{-}/{\rm He^{2+}}$ ratio \cite{e_he_ratio}
and $e^{-}/p^{+}$ ratio \cite{e_p_ratio} variations over 
a long period appear to support this charge dependence. 
Since protons and antiprotons are different only in
their charge sign, the simultaneous measurements of both particles are most
suitable for studying the charge dependence of solar modulation.
 
We report here a new measurement of cosmic-ray $\bar{p}$ spectrum 
in the energy range between 0.18 and 4.20 GeV, based on 384 events  
detected by the BESS spectrometer in 1998 flight.
In total, 848 $\bar{p}$ events has been accumulated
from the measurement in successful four flights in 1993, 1995, 1997
and 1998.
The spectrum measured in 1998 is compared with the previous results 
\cite{BESS97,YOMO,IMAX,CAPRICE,MA98,MASS91}, in order to observe changes 
with solar activity.

\section{BESS Detector}

\begin{figure}[p]
\centerline{\epsfxsize=13cm \epsffile{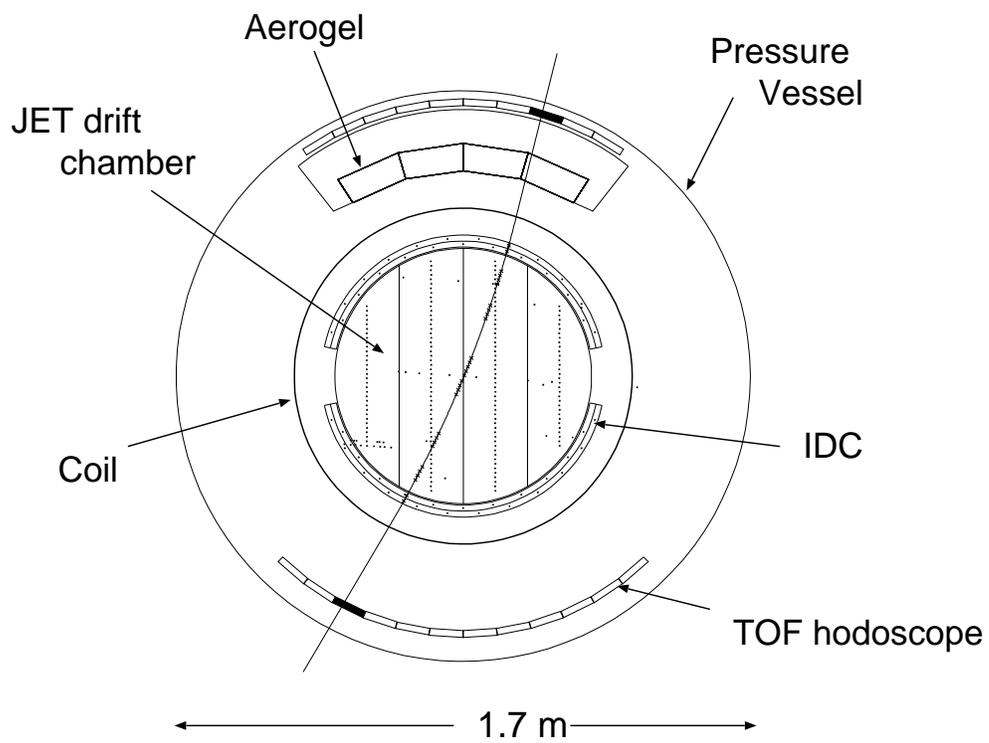}}
\caption{
Cross-sectional view of the BESS 1998 detector 
with one of the $\bar{p}$ events.
}
\label{fig:detector}
\end{figure}

The BESS detector, shown in Fig.\ref{fig:detector},
was proposed and developed
as a high-resolution spectrometer to perform
searches for rare cosmic-rays as well as
precision measurements on absolute fluxes of various cosmic-ray particles
\cite{OR87,YAM,DETEC}.
Various new detector technologies developed for collider
experiments were incorporated to the spectrometer. 
A uniform field of 1 Tesla was produced by a thin
(4 g/cm$^2$) superconducting coil \cite{YA88}, through
which particles could pass with small interaction probability.  The
magnetic-field region was filled with central tracking devices.  
This configuration achieved a geometrical acceptance of
0.3 m$^2$Sr, and it was an order of magnitude larger than those of
previous cosmic-ray spectrometers for balloon experiments.
The $r\phi$ tracking in the central region
was performed by fitting up to 28 hit-points, each with 200 $\mu$m resolution,
in a jet-type drift (JET) chamber  and two inner drift chambers (IDC's)
resulting in a magnetic-rigidity ($R \equiv Pc/Ze$) resolution of
0.5 \% at 1 GV/$c$.
Tracking in the $z$ coordinate was done by
fitting points in IDC's measured by vernier pads 
with an accuracy of 470 $\mu$m 
and points in the JET chamber measured 
by charge division with an accuracy of 2.5 cm.
The continuous and redundant 3-dimensional tracking enabled us
to recognize multi-track events and tracks having interactions
or scatterings, thus minimizing such backgrounds. The upper and
lower scintillator hodoscopes \cite{SIK00}
measured two independent $dE/dx$ and the
time-of-flight (TOF) of particles.  In addition, $dE/dx$ in the
drift chamber gas was obtained as a truncated mean of the
integrated charges of the hit-pulses. 
The scintillator hodoscopes consisted of ten upper and twelve lower
plastic scintillators 
and photo-multiplier tubes (PMT's) attached at each end of scintillators.
The hodoscopes were placed at outer-most radii and
the timing resolution of each hodoscope was 55 psec rms,
resulting in $\beta^{-1}$ resolution of 0.014, where $\beta$ is 
defined as particle velocity divided by the speed of light.
Furthermore a Cherenkov counter \cite{AS98} 
with silica-aerogel radiator \cite{SM98}
was installed below the upper TOF hodoscopes.
We selected the radiator having a refractive index of 1.020 
(while 1.032 for BESS 1997), in order to veto $e^{-}/\mu^{-}$ backgrounds
up to 4.2 GeV.

In 1998 flight, the payload was launched on July 29 from Lynn Lake, Manitoba
to Peace River, Albata in northern Canada, where the cut-off rigidity is
less then 0.5 GV/$c$.
The scientific data were taken for a live time of 60462 sec
at altitudes ranging from 37.4 to 35.7 km
(with residual air of 4.8--5.3 g/cm$^2$).
All the detector components worked well during the flight and
19 million cosmic-ray events were recorded onto two magnetic tapes
(the total recorded data size was 33.5 Gbytes).
The first-level trigger
was provided by a coincidence between the top and the bottom
scintillators, with the threshold set at 1/3 of the pulse height
from minimum ionizing particles.  The second-level trigger, which
utilized the hit-patterns of the scintillator-hodoscopes and
the inner drift chambers (IDC), first rejected unambiguous null- and
multi-track events and then made a rough rigidity-determination
to select negatively-charged particles predominantly.
In addition, one of every 60
first-level triggered events were recorded
in order to build a sample of unbiased triggers.

\section{Analysis}

Since the BESS instrument had highly symmetrical detector-configuration,
we assumed that $\bar{p}$'s behaved similarly as protons 
in the instrument except for deflection in the magnetic field
and the inelastic interactions.
Thus, all selection criteria were defined
based on the measured properties of protons.
At first, we selected events
with (i) a single down-going particle fully contained
in the fiducial region of the tracking volume,
(ii) only one hit in the upper TOF hodoscope
and (iii) one or two hits in the lower TOF hodoscope.
Then the following cuts were applied in order to ensure correct
measurements.
(1) The numbers of used hits in the $r\phi$-fitting 
and in the $z$-fitting should be more than 10 and 6, respectively:
(2) The reduced $\chi^{2}$ of the fitted $r\phi$ and $z$ track had to
be less than 6.5 and 6.0, respectively:
(3) Since signals from some PMT's were somewhat noisy,
we rejected wrong hits in the TOF hodoscopes for better TOF measurement:
(4) The $z$ position determined from the left-right time difference measured
by the PMT's should match the $z$-impact point of the
extrapolated track at the TOF hodoscopes:
(5) The ratio of the signal amplitude of the left and right PMTs should be
consistent with the $z$-impact point of the extrapolated track:
The instrument was suspended with the balloon via a
parachute (50m H $\times$ 1m $\phi$) of 340 kg, situated 24 meter above 
the instrument.
Particles passing through the parachute would interact and change
their energy. Therefore,
(6) the extrapolated track should not pass through the parachute:
(7) The three d$E$/d$x$ measurements in the scintillator-hodoscopes 
and JET chamber were loosely required as functions of $R$
to be compatible with proton or $\bar{p}$:
(8) The proper value of 1/$\beta$ was required as a function of $R$:
The combined efficiency of these off-line selections
was 71 -- 75 \% for $R$ from 0.5 to 5 GV/$c$.

These simple and highly-efficient selections were sufficient for a very
clean detection of $\bar{p}$'s in the low velocity ($\beta < 1/1.15$) region.
At higher velocities, the $e^{-}/\mu^{-}$ background
started to contaminate the $\bar{p}$ band, where
we required the Cherenkov veto; i.e.,
(1) the particle trajectory to traverse the fiducial volume of the aerogel,
and (2) the Cherenkov output
to be less than 0.09 of the mean output from $e^{-}$.
This cut reduced the acceptance by 20 \%,
but rejected $e^{-}/\mu^{-}$ backgrounds by a factor of 4400, while
keeping 92 \% efficiency for protons and $\bar{p}$'s
which crossed the aerogel with rigidity below the threshold
(4.7 GV/$c$).

\begin{figure}[hbtp]
\centerline{\epsfxsize=13cm \epsffile{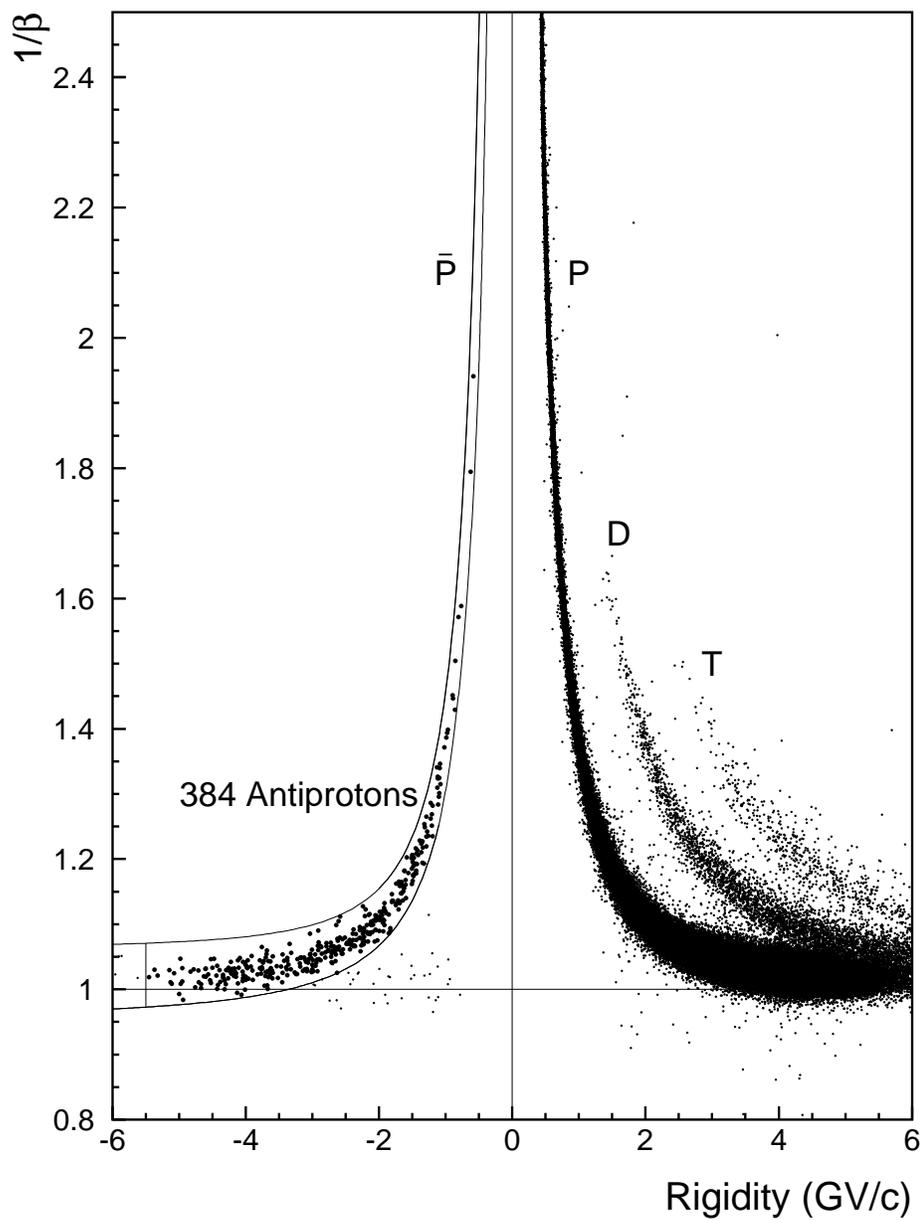}}
\caption{
The identification of $\bar{p}$ events.
The solid curves define the $\beta^{-1}$--$R$ region
 and the $\bar{p}$ mass band used for the spectrum measurement.
}
\label{fig:id98}
\end{figure}

Fig.\ref{fig:id98} shows the $\beta^{-1}$ versus $R$ plot
for the surviving events.  We see a clean narrow band of 384 $\bar{p}$'s
at the exact mirror position of the protons.  The $\bar{p}$ band
was slightly contaminated with the $e^{-}/\mu^{-}$ backgrounds
due to the inefficiency of the aerogel Cherenkov counter,
and in the flux calculation 
we subtracted this background whose amount was estimated as follows:
(1) We counted the number of $e^{-}/\mu^{-}$ events overlapping
with the $\bar{p}$ band before the requirement on the Cherenkov output.
(2) We obtained the probability that $e^{-}/\mu^{-}$ gave a lower
Cherenkov output than the threshold, from the distribution of 
the output for high rigidity protons 
($\geq 25$GV) which should emit enough output.
(3) From the above two quantities we estimated the amount 
of the $e^{-}/\mu^{-}$ background as $0$ \%, $0.6$ \% and $4.2$ \%
at 0.25, 2 and 4 GeV, respectively.
Backgrounds of albedo and of mis-measured
positive-rigidity particles were totally excluded
by the excellent $\beta^{-1}$ and $R^{-1}$ resolutions.
To check against the ``re-entrant albedo''
background, we confirmed that the trajectories of all
$\bar{p}$'s can be traced numerically through the Earth's
geomagnetic field back to the outside of the geomagnetic sphere.

We obtained the $\bar{p}$ energy spectrum at the top of the atmosphere
(TOA) in the following way:
The geometrical acceptance of the spectrometer was
calculated by a Monte Carlo simulation \cite{MATSU}, 
which was consistent with an
analytical calculation \cite{GEO}.
The fraction of the live data-taking time was directly measured 
as 86.4 \% by counting 1 MHz clock.
The efficiencies of trigger and of the
off-line selections were determined by using protons 
in the unbiased trigger samples.
The TOA energy of each event was
calculated by tracing back the particle through the detector material
and the air.  The survival probability of the $\bar{p}$'s through
the air and instrument was evaluated 
by {\sc geant/gheisha} simulation \cite{MATSU}, which incorporated
detailed material distribution, realistic detector performance
and correct $\bar{p}$-nuclei cross sections \cite{CRSEC}.  
The $e^{-}/\mu^{-}$ background was subtracted, 
whose amount is described above. And we subtracted
the expected number of atmospheric $\bar{p}$'s, produced by the
collisions of cosmic rays in the air.  
The subtraction amounted to
$17\pm3$ \%, $21\pm4$ \% and $22\pm6$ \% at 0.3, 0.7 and 2 GeV, 
respectively, where the errors correspond to the maximum difference
among the three independent recent calculations \cite{MITSUI,PF96,ST96}.
Table \ref{tab:pbsummary} contains resultant 1998 $\bar{p}$ fluxes 
and $\bar{p}/p$ flux ratios at TOA.
The first and the second error 
represent the statistical \cite{FEL98} and systematic errors, respectively.
The dominant systematic error at low energies is the uncertainty in the
evaluation of the interaction losses, to which we attribute $\pm$ 15 \%
relative error. At high energies, the uncertainty in the atmospheric
$\bar{p}$ calculation becomes dominant.
The statistical errors are always larger than the systematic errors
by a large factor especially at the low energies.

\section{Result and Conclusion}

\begin{figure}[hbtp]
\centerline{\epsfxsize=13.0cm \epsffile{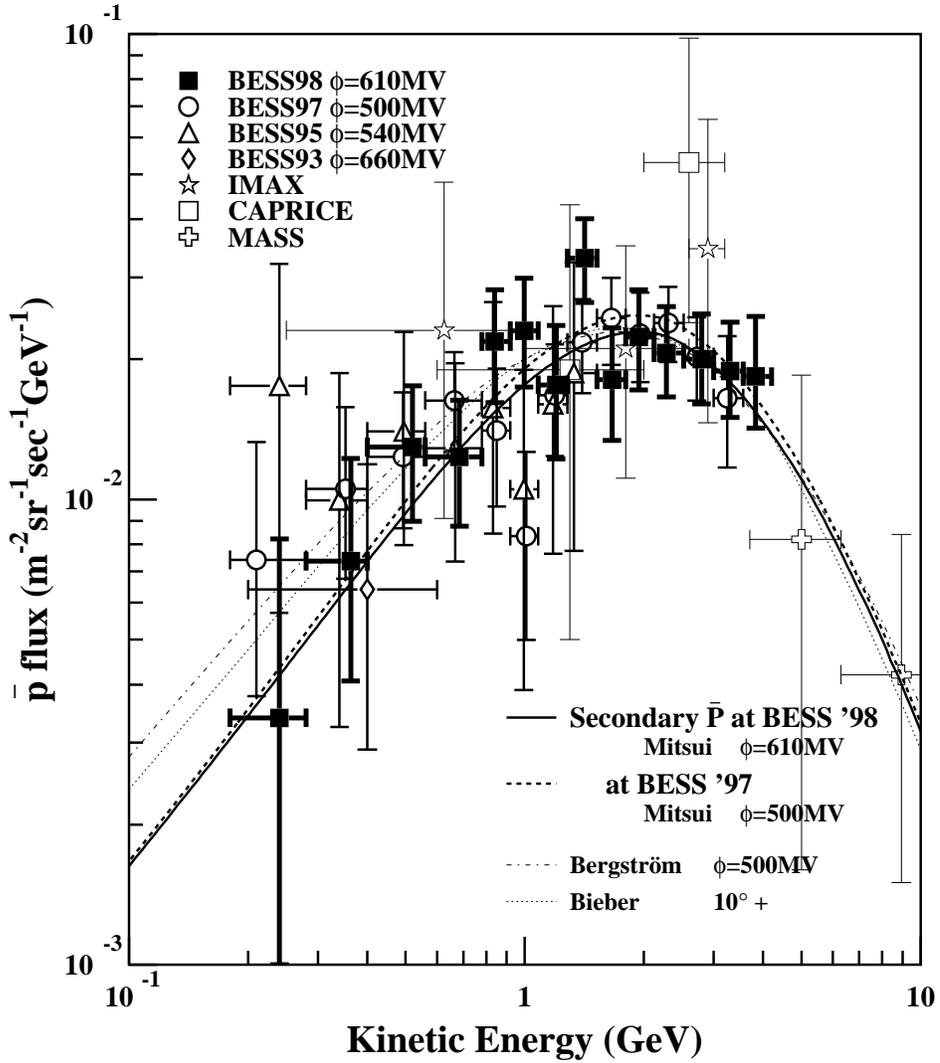}}
\caption{
The BESS 1998 antiproton spectrum at the top of the atmosphere,
together with the previous data.
The thick solid curve represents the expected spectrum 
for secondary $\bar{p}$ \cite{FRB} at the 1998 flight ($\phi_{98}$=610MV)
and well reproduces the BESS 1998 spectrum in the whole energy region.
The expected spectrum at the 1997 flight ($\phi_{97}$=500MV) was
also calculated in the same way as the 1998, 
represented by the thick dashed curve.
Although the curve agrees with the BESS 1997 spectrum in the
peak region, the agreement is less consistent in the low energy region.
Also shown are other calculations \cite{BERG,BIEB}
for secondary $\bar{p}$
at solar minimum (corresponding to the solar activity level 
at the BESS 1997 flight).
}
\label{fig:pbflux}
\end{figure}

Fig.\ref{fig:pbflux} shows the BESS 1998 $\bar{p}$ spectrum,
together with the previous measurements 
\cite{BESS97,YOMO,IMAX,CAPRICE,MA98,MASS91}.
By using 1998 flight data, 
we have measured the $\bar{p}$ spectrum in a wider energy region 
of 0.18 to 4.20 GeV and
detected again a clear peak around 2 GeV as we did in 1997.
As recently pointed out
\cite{MI96}, increase of solar activity suppresses
any soft primary $\bar{p}$ component while only modestly affects
the secondary $\bar{p}$ spectrum.
The solar activity at the time of the BESS 1998 flight was greater than
the 1995 and 1997 flights \cite{NEUT}.
Therefore, the shape of the 1998 spectrum should be closer to 
the shape of the spectrum composed only of secondary $\bar{p}$,
even if the primary $\bar{p}$'s existed.
Shown also in Fig.\ref{fig:pbflux} is a recent theoretical 
calculation \cite{FRB}
for the secondary $\bar{p}$ at the BESS 1998 flight
(solar modulation parameter $\phi_{98}$=610MV, 
which was determined from BESS 1998 proton spectrum \cite{SAN98}).
In the whole energy region, 
this calculation agrees well with our spectrum
at 86\% confidence level.
This implies that secondary $\bar{p}$ is
the dominant component of the cosmic-ray $\bar{p}$'s
and the model used for the calculation is
basically correct. The expected spectrum at BESS
1997 was also calculated in the same way by
using the solar modulation parameter  $\phi_{97}$=500MV
as represented by the dashed curve in Fig.\ref{fig:pbflux}.
It appears to be insensitive to the change of the modulation parameter.
This is because the interstellar $\bar{p}$ flux steeply drops toward 
the low energy in contrast to the interstellar proton spectrum \cite{LAB97}.
In fact, the measured spectra are nearly constant in the peak region.
However, in the low energy region (0.18 -- 0.78 GeV),
the agreement between the 1997 spectrum and the calculation
is less consistent.
Moreover, the 1995 spectrum shows larger deviation from the calculation
in spite of similar modulation parameter ($\phi_{95}$=540MV) as the BESS 1997.
This might be due to statistical fluctuation,
or might suggest
a contribution of low-energy primary $\bar{p}$ from novel source such as
annihilation of neutralinos \cite{BERG,SILK,MI96} or 
the evaporating primordial black holes \cite{HA75,MA96}.
Because $\bar{p}$'s from these ``primary'' sources, if they exist,
are expected to be prominent at low energies and 
to exhibit large solar modulations \cite{MI96}.

\begin{figure}[hbtp]
\centerline{\epsfxsize=13.0cm \epsffile{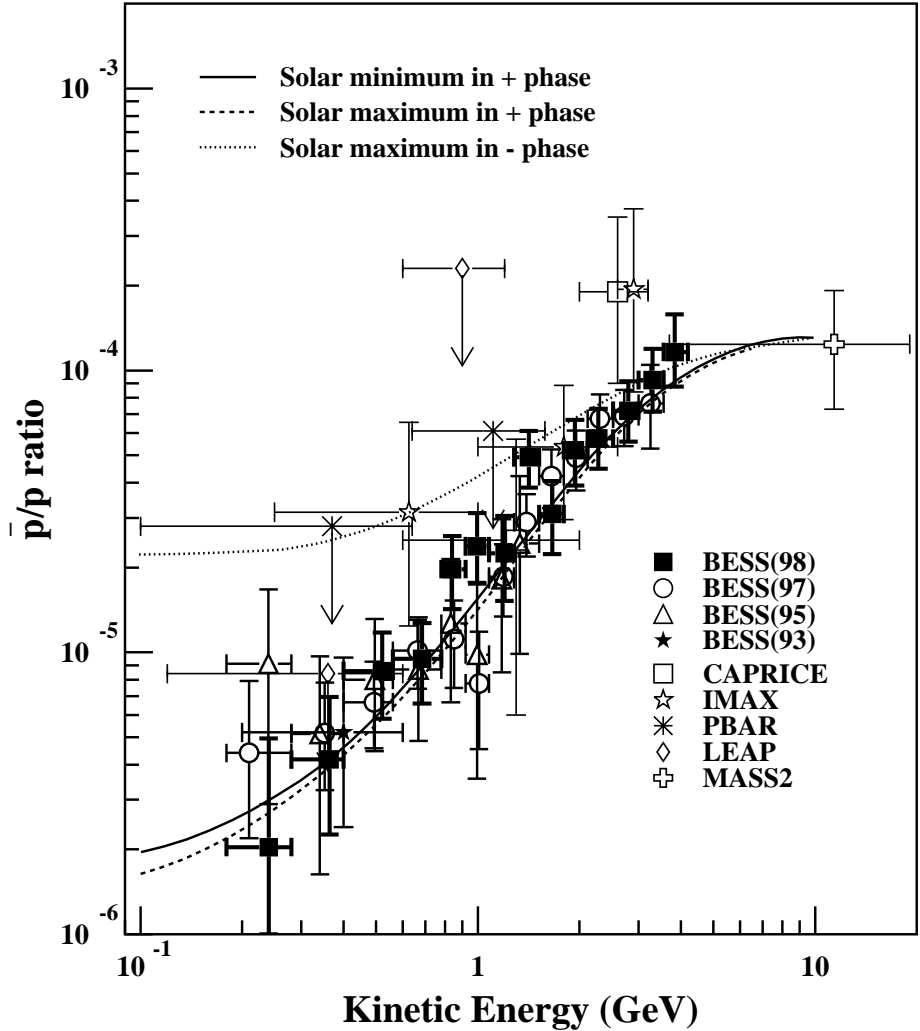}}
\caption{
Comparison of the BESS 1993, 1995, 1997 and 1998 $\bar{p}/p$ ratios 
with previous data \cite{CAPRICE},
and the calculation \cite{BIEB} taking the charge dependence 
of the solar modulation into account. 
The solid and dashed curves represent the expected 
$\bar{p}/p$ ratio at the solar minimum and at the solar maximum in
the positive Sun's polarity.
When the polarity switches, the ratio is expected to increase as
represented by the dotted curve.
}
\label{fig:ppratio}
\end{figure}

\begin{figure}[hbtp]
\centerline{\epsfxsize=13.0cm \epsffile{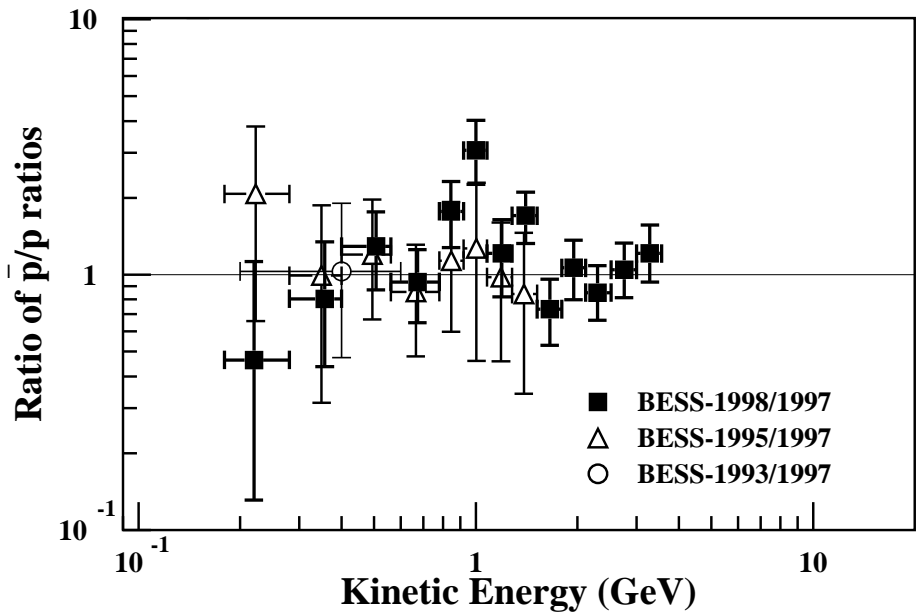}}
\caption{
The ratios of the $\bar{p}/p$ ratio normalized by the BESS 1997 data
are distributed around the unity within the statistical fluctuation.
The BESS 1993, 1995, 1997 and 1998 data were 
measured in the positive polarity phase.
Therefore, it is shown that the $\bar{p}/p$ ratio is nearly identical during 
the positive polarity phase.
}
\label{fig:ratio}
\end{figure}

In order to study the solar modulation, the $\bar{p}/p$ ratio is more adequate,
which is expected to clarify the charge-dependence 
of the solar modulation \cite{BIEB}.
In Fig. \ref{fig:ppratio} the BESS 1998 $\bar{p}/p$ ratio shows 
almost similar spectrum as the spectrum at the solar minimum
(the 1997 data), especially around 2 GeV where we have better statistics.
The lowest energy point of our data might suffer some uncertainties
in the proton flux due to large subtraction of atmospheric secondary protons,
which is significant in the low energy region \cite{SAN98}.
Despite the general advantage of taking $\bar{p}/p$ ratio to cancel
various systematic errors, such as live-time and geometrical acceptance,
it should be noted that the most significant systematic errors in the analysis
come from the uncertainties in the atmospheric $\bar{p}$ calculation 
and in the interaction losses,
which can not be canceled even in the form of the ratio.
Taking the charge-dependence of the solar modulation into account \cite{BIEB},
(i) the $\bar{p}/p$ ratio is expected to be nearly identical
during the positive Sun's polarity phase;
and (ii) when the polarity switches into the negative,
the $\bar{p}/p$ ratio shall rapidly increase as represented 
by the dotted curve in Fig. \ref{fig:ppratio}.
The BESS 1993, 1995, 1997 and 1998 spectra, 
measured in the positive polarity
phase, were nearly identical as shown in Fig. \ref{fig:ppratio}
and consistent with the prediction.
It can be more clearly understood in the form
of the ratios of $\bar{p}/p$ ratio as shown in Fig. \ref{fig:ratio}.
The ratios of $\bar{p}/p$ ratio normalized by the BESS 1997 data
are consistently distributed around the unity
within the statistical fluctuation.

Our measurements on the $\bar{p}/p$ ratio are consistent with 
the prediction taking the charge
dependent solar modulation into account.
However, it is too early to conclude that the evidence for 
the charge dependence is exhibited.
We are planning further annual flights in the following 
solar-maximum period in which the polarity switches into the negative.
The observation of the increase would be a definite evidence 
for the importance of the charge dependent effect in the solar modulation.
At the same time, measurements of the $\bar{p}$ spectrum
through the solar-maximum period itself have a crucial importance
to determine the origin of the low energy  $\bar{p}$'s.
As previously mentioned, any soft $\bar{p}$ component 
from the primary sources would 
be suppressed as the solar activity increases, while 
the secondary $\bar{p}$ spectrum is affected modestly.
Therefore, to observe variations in the low energy region 
would help us to draw a firm conclusion on whether
we are seeing the primary $\bar{p}$'s from novel sources.

Sincere thanks are given to NASA/GSFC/WFF balloon office 
and NSBF for balloon launch and recovery of the payload
in Canada campaign.
The authors deeply thank ISAS and KEK for their continuous
support and encouragement  for the BESS experiment. They would specially
thank Prof. T. Sumiyoshi, Dr. R. Suda and Mr. T. Ooba 
for their kind guidance
and cooperation to develop Aerogel Cherenkov Counters for
the flight in 1998.
The analysis was performed by using the computing facilities
at ICEPP, Univ. of Tokyo.  This experiment was supported by
Grant-in-Aid for Scientific Research in Monbusho,
Heiwa Nakajima Foundation in Japan, and by NASA in the USA.

\begin{table}[hbtp]
\caption{
Antiproton fluxes 
(in $\times 10^{-2}$ m$^{-2}$s$^{-1}$sr$^{-1}$GeV$^{-1}$)
and $\bar{p}/p$ ratios (in $\times 10^{-5}$) at TOA.
T (in GeV) define the the kinetic energy bins.
$\rm{N}_{\bar{p}}$ and $\overline{\rm{T}}_{\bar{p}}$,
respectively, are the number of observed antiprotons and their mean kinetic energy
in each bin.
}
\label{tab:pbsummary}
\renewcommand{\arraystretch}{2.00}
\begin{tabular}{c|cccc}
\hline
 & \multicolumn{4}{c}{BESS 1998} \\
T (GeV)
 & $\rm{N}_{\bar{p}}$ 
 & $\overline{\rm{T}}_{\bar{p}}$
 & $\bar{p}$ flux
 & $\bar{p}/p$ ratio\\
\hline
0.18 - 0.28
 &     2
 &   0.24
 & $0.34^{+0.48\,+0.06}_{-0.23\,-0.06} $
 & $0.20^{+0.29\,+0.04}_{-0.14\,-0.04} $ \\ 
0.28 - 0.40
 &     6
 &   0.36
 & $0.74^{+0.48\,+0.09}_{-0.32\,-0.09} $
 & $0.42^{+0.27\,+0.06}_{-0.18\,-0.06} $ \\ 
0.40 - 0.56
 &    15
 &   0.52
 & $1.30^{+0.44\,+0.14}_{-0.37\,-0.14} $
 & $0.86^{+0.29\,+0.13}_{-0.25\,-0.13} $ \\ 
0.56 - 0.78
 &    22
 &   0.68
 & $1.23^{+0.38\,+0.13}_{-0.33\,-0.13} $
 & $0.95^{+0.29\,+0.14}_{-0.26\,-0.14} $ \\ 
0.78 - 0.92
 &    23
 &   0.84
 & $2.19^{+0.60\,+0.21}_{-0.53\,-0.21} $
 & $1.97^{+0.54\,+0.27}_{-0.48\,-0.27} $ \\ 
0.92 - 1.08
 &    25
 &   1.00
 & $2.30^{+0.65\,+0.21}_{-0.52\,-0.21} $
 & $2.38^{+0.67\,+0.32}_{-0.54\,-0.32} $ \\ 
1.08 - 1.28
 &    21
 &   1.20
 & $1.76^{+0.58\,+0.16}_{-0.52\,-0.16} $
 & $2.25^{+0.74\,+0.31}_{-0.66\,-0.31} $ \\ 
1.28 - 1.52
 &    41
 &   1.42
 & $3.30^{+0.65\,+0.29}_{-0.58\,-0.29} $
 & $4.93^{+0.97\,+0.65}_{-0.87\,-0.65} $ \\ 
1.52 - 1.80
 &    29
 &   1.66
 & $1.81^{+0.49\,+0.20}_{-0.43\,-0.20} $
 & $3.09^{+0.84\,+0.46}_{-0.73\,-0.46} $ \\ 
1.80 - 2.12
 &    37
 &   1.94
 & $2.23^{+0.54\,+0.23}_{-0.46\,-0.23} $
 & $5.22^{+1.26\,+0.74}_{-1.08\,-0.74} $ \\ 
2.12 - 2.52
 &    42
 &   2.28
 & $2.07^{+0.50\,+0.18}_{-0.36\,-0.18} $
 & $5.74^{+1.38\,+0.77}_{-1.00\,-0.77} $ \\ 
2.52 - 3.00
 &    46
 &   2.80
 & $2.00^{+0.45\,+0.22}_{-0.33\,-0.22} $
 & $7.20^{+1.63\,+1.06}_{-1.19\,-1.06} $ \\ 
3.00 - 3.56
 &    46
 &   3.30
 & $1.89^{+0.45\,+0.25}_{-0.29\,-0.25} $
 & $9.24^{+2.21\,+1.54}_{-1.44\,-1.54} $ \\ 
3.56 - 4.20
 &    29
 &   3.83
 & $1.84^{+0.58\,+0.26}_{-0.32\,-0.26} $
 & $11.63^{+3.66\,+2.01}_{-2.05\,-2.01} $ \\ 
\hline
\end{tabular}
\end{table}

\renewcommand{\arraystretch}{1}

\end{document}